\documentclass[twocolumn]{article}
\usepackage{graphicx}

\begin{document}

\title{Teaching Waves with Google Earth}

\author{Fabrizio Logiurato \\ \small{Dipartimento di Fisica, Universit\`a di Trento}
\\ \small{INO-CNR BEC Center and Physics Department, Trento University, I-38123 Povo, Italy}}

\date{31nd August 2011}
\maketitle 
\begin{abstract}
Google Earth is a huge source of interesting illustrations of various natural phenomena. 
It can represent a valuable tool for science education, not only for teaching geography and geology, 
but also physics. Here we suggest that Google Earth can be used for introducing in an attractive way 
the physics of waves.  
\end{abstract}

\section{Introduction}
The physics of water waves is in general very complicated. Water waves are usually used as an 
example of waves in elementary courses, but as Feynman said "they are the worst possible example,
 because they are in no respect like sound and light; they have all the complications that waves can have" \cite{Feynman}.  
 
For instance, since water is practically incompressible with the pressure in play, a wave on the 
surface is not purely transverse: as the wave moves forward, water must move away from the trough 
 to the crest, and the particles of water near the surface move approximately in circles. 
 
Also establishing the velocity of a water wave is not an easy theoretical question \cite{Barber}, \cite{Bascom}. 
 If the depth of  water is large, (more than a wavelength), and the wavelength is long 
 (more than a meter), the phase velocity of an approximately sinusoidal wave is proportional to the 
 square root of the wavelength $\lambda$ :

\begin{equation} \label{Gw1}
v_{phase}=\sqrt{\frac{g \lambda}{2 \pi}}  \qquad {\rm Gravity \, waves} \,\,   {\rm (deep \, water)} \,,
\end{equation}

\noindent
where $g$ is the gravitational acceleration.  A wave with a longer wavelength goes faster than 
one with a shorter wavelength. 

For instance, the wind of a storm in the open sea produces waves of 
all  lengths:  waves with a long wavelength reach the beach first, waves with a shorter wavelength 
are slower and they arrive later.  The same thing happens if it is a boat that makes waves: 
 long waves of its wake arrive at the beach followed by shorter waves.
 
If waves are very short, like ripples in a ripple-tank, the main force on them  
is not the gravitation but the surface tension. For such waves, denominated capillary waves, 
the phase velocity is

\begin{equation} \label{Cw1}
v_{phase}=\sqrt{\frac{ 2 \pi T}{\lambda \rho}}  \qquad {\rm Capillary \,\, waves} \,,
\end{equation}


\begin{figure*}[!ht]
\centering
\includegraphics[width=16.5cm]{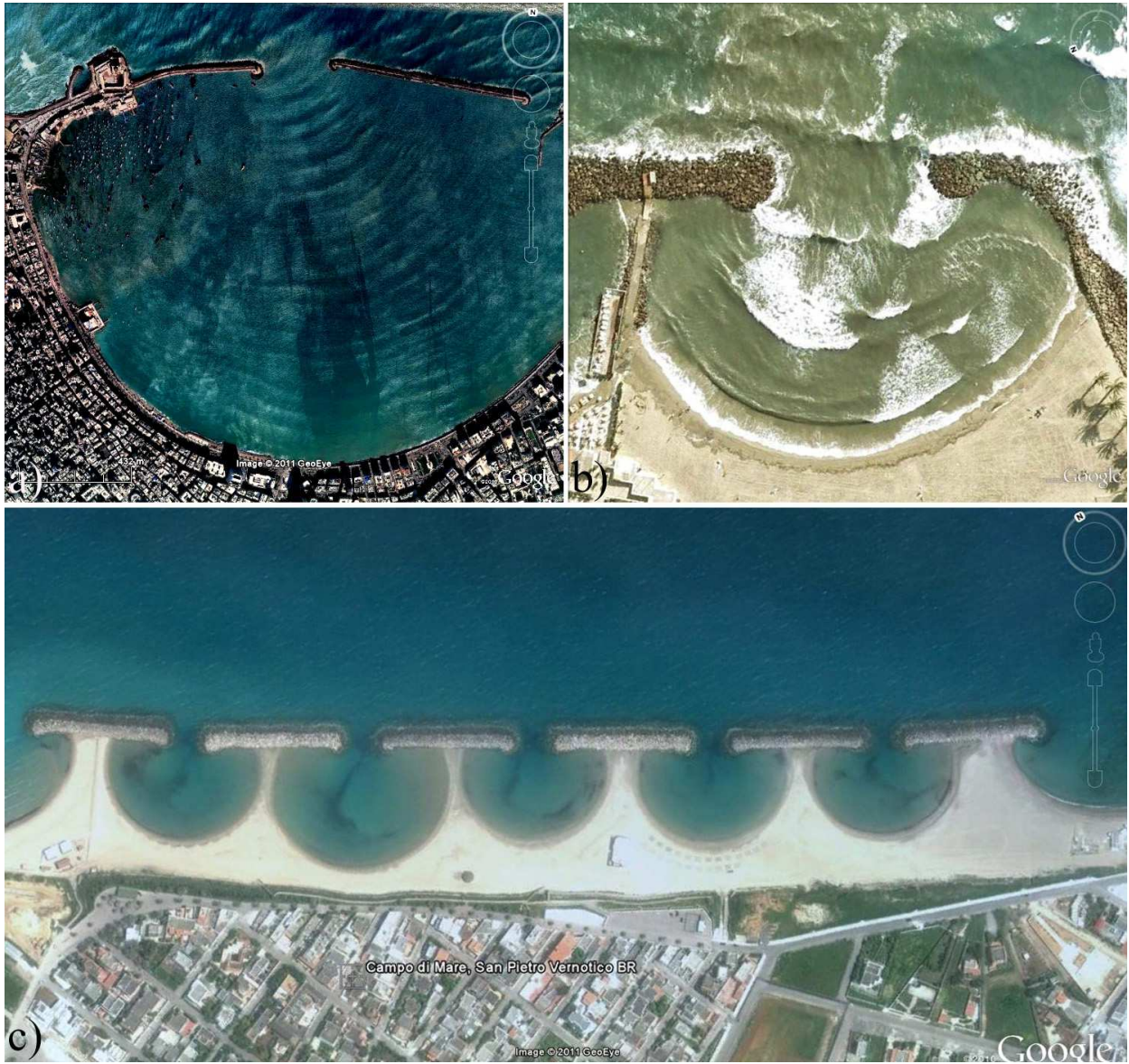}
\caption{Wave diffraction through an opening: a) Alexandria of Egypt, 12/14/2010, coordinates: $31^0 \, 12' \,28.56" \,N$, $29^0\, 53'\, 34.66" \,E$; b)
Th\'eoule-sur-Mer,France,10/26/2006, coordinates: $43^0 \, 31' \, 54.86" \,N$, $6^0\, 56'\, 59.41" \,E$. c)   Wave diffraction causes circular erosion of the beach: Campo di Mare, Italy, 4/18/2010,  coordinates: $40^0 \,32' \, 27.33" \, N$, $18^0 \, 04' \, 09.17" \, E$.}
\end{figure*}



\begin{figure*}[!ht]
\centering
\includegraphics[width=16.5cm]{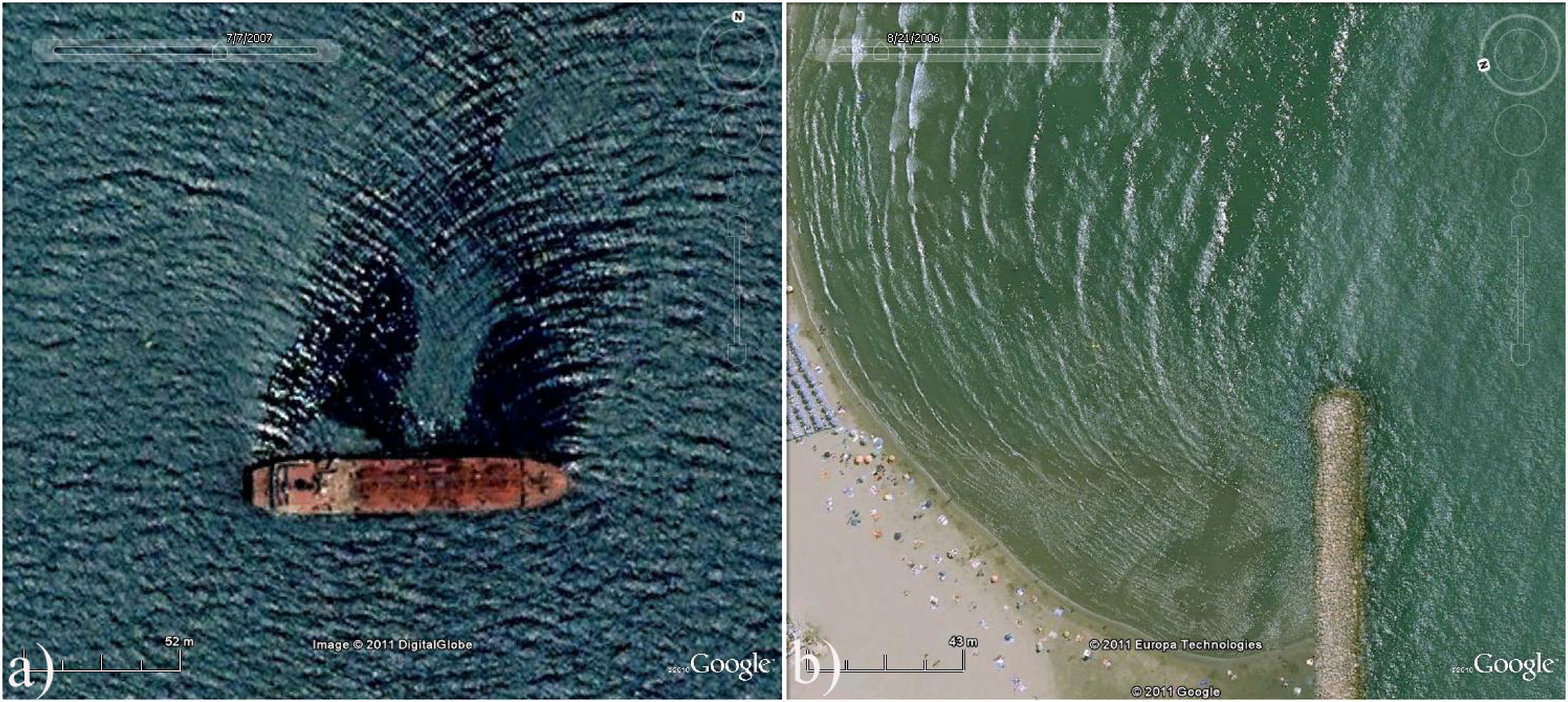}
\caption{
a) Diffraction produced by a boat: Cyprus, 7/7/2007, coordinates: $34^0 \, 56'\, 27.21" N$,  $33^0 \, 39'\, 17.36" E$.  
b) Diffraction of waves against the end of a protection barrier: La Grande-Motte, France, 8/21/2006, coordinates: $43^0 \, 33' \, 18.71" N$,  
$4^0 \, 05'\, 20.01" E$.}
\end{figure*}


\noindent
where $T$  is the surface tension and $\rho$   the density of  water. So, for this kind of waves, 
velocity is higher when the wavelength becomes shorter.

Capillary waves in water have wavelength of a few cm or less. 
In general, we have both actions of gravity and surface tension, 
so the correct formula is more complicated and it has to contain both effects.

The group velocity can be deduced from the previous relations: it follows that 
for gravity waves the group velocity is less than the phase velocity, for ripples the group velocity
 is higher than the phase velocity. This is responsible for the different forms of wakes, such as those 
 caused by a boat or a stick moving in water \cite{Lighthill}.
 
If the water is very shallow, when the depth $d$ is much less than a wavelength, we have to consider
the friction effect of the sea floor. Then another approximate formula for the velocity must be used for gravity waves:
 
\begin{equation} \label{Gws}
v_{phase}=\sqrt{g d} \quad {\rm Gravity \, waves \, (shallow \, water)} \,.
\end{equation}

\noindent
Therefore,  in shallow water, all waves travel with the same speed, which depends only on  the depth of water. 
When the water depth decreases approaching to the shore, waves reduce their velocity. 

In spite of the difficulty of the matter, many phenomena occurring on water surfaces can be qualitatively 
described with the same principles of wave physics taught at school  and with just a few concepts introduced above. 

As already suggested \cite{Ryder} - \cite{Logiuratob}, Google Earth can be of valuable help in teaching physics.
 Young pupils usually find the images of wave phenomena on sea, lakes or rivers, much more interesting than usual 
 pictures of the books, normally taken from experiments with the ripple tank, or just drawn. 
For this reason, we think that Google Earth pictures can help to introduce in an effective way some fundamental 
concepts of wave physics \cite{Google}. For instance, we can find a lot of beautiful illustrations of phenomena as reflection, 
refraction, diffraction and interference. 
\section{A Few Examples}

According to Huygens' Principle \cite{French}, a wave front which hits the points of a medium makes these points centers of new disturbances 
and sources of secondary waves. The amplitude of a secondary wave is at its maximum in the direction of propagation and decreases to 
zero towards the perpendicular direction. 
We can imagine a wave front constituted by many Huygens' secondary waves. When a point of the front hits an obstacle or another point 
of the medium, that point becomes the source of another secondary circular wave. So we can describe the spreading out of waves when they 
go through an opening or why they can overshoot an obstacle. 
We observe these diffraction phenomena when the wavelength is comparable with the dimensions of the obstacle.


\begin{figure*}[!ht]
\centering
\includegraphics[width=16.5cm]{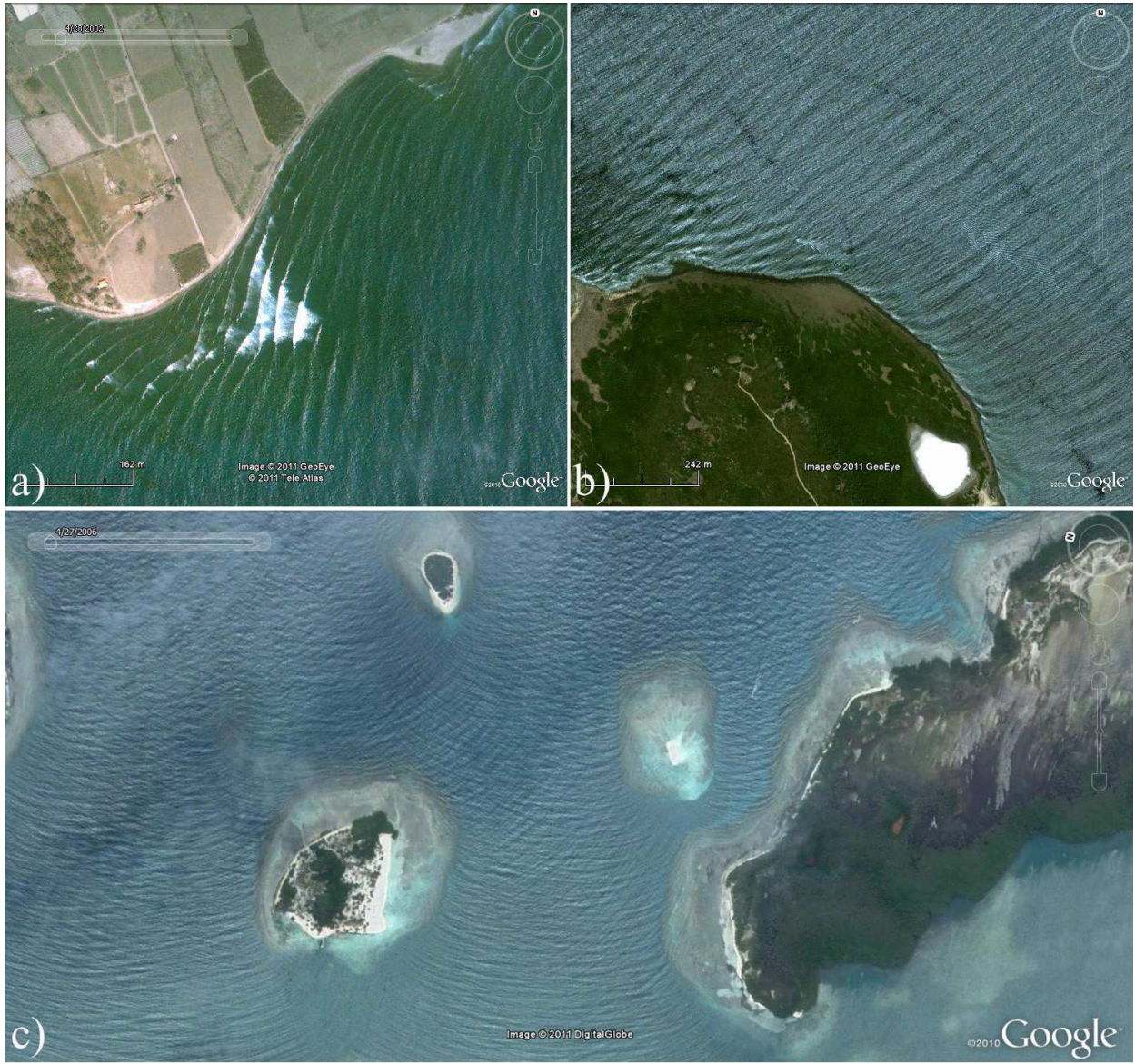}
\caption{
Examples of wave refraction, the wave fronts bend approaching the beach a) Villaricos, Spain, 4/26/2002, coordinates:   
$37^0 \, 14' \, 04.59" \,N$, $1^0\, 47'\, 04.79" \,E$.
b) Sardegna, Italy, 8/18/2010, coordinates: 
$39^0 \, 46' \, 12.64" \,N$, $8^0\, 27'\, 28.82" \,E$.
c) Chichiriviche, Venezuela, 4/27/2006, coordinates:
$10^0 \, 55' \, 29.88" \,N$, $68^0\, 15'\, 26.93" \,W$.}
\end{figure*}



\begin{figure*}[!ht]
\centering
\includegraphics[width=16.5cm]{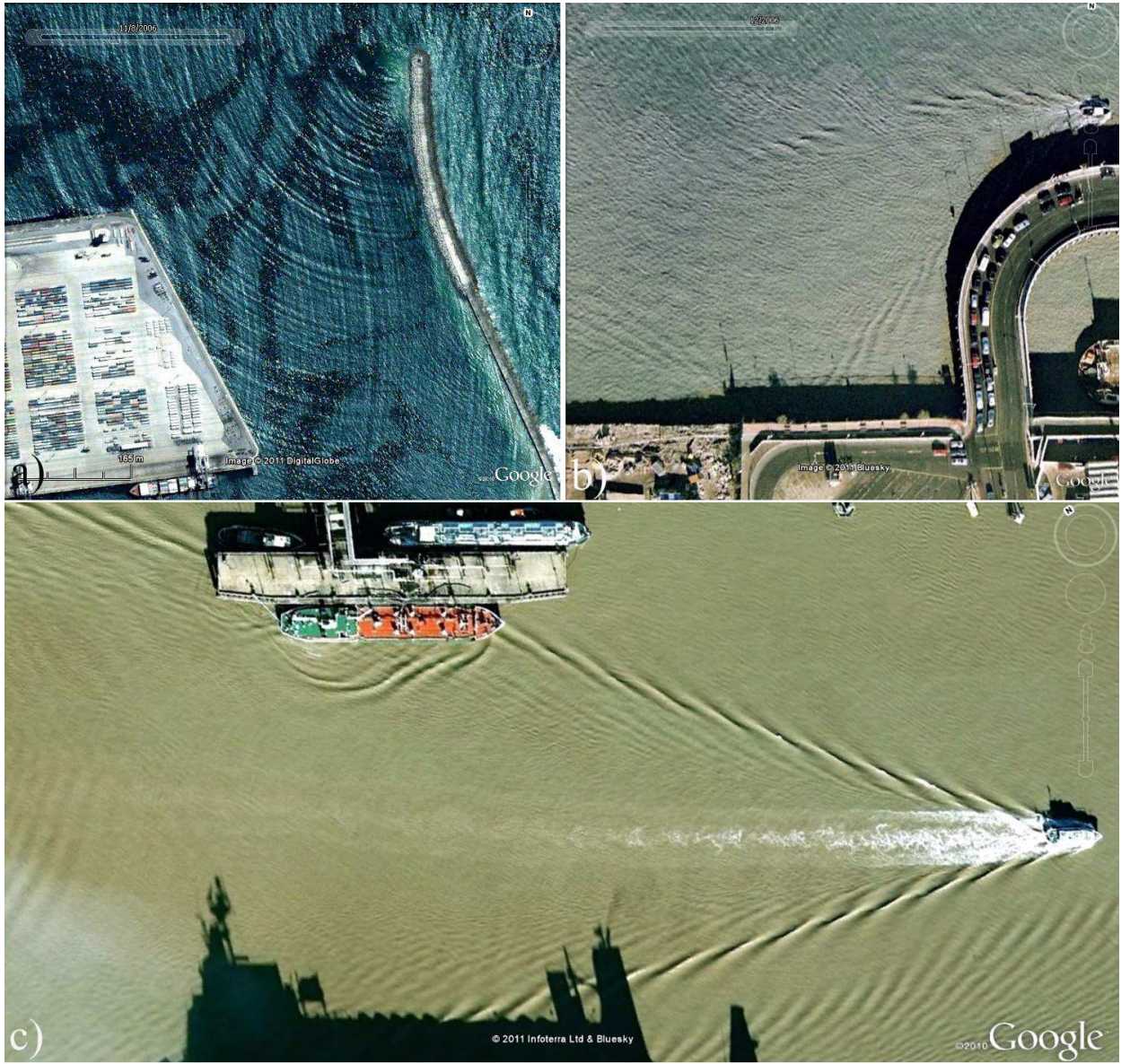}
\caption{
Examples of diffraction and reflection of circular waves: a) Port Elizabeth, South Africa 11/8/2006, coordinates:   
$33^0 \, 57' \, 19.98" \,S$, $25^0\, 38'\, 33.05" \,E$. Reflections of a boat wake by an obstacle: 
b) River Thames, London, England, 11/06/2006, coordinates: 
$51^0 \, 29' \, 42.06" \,N$, $0^0\, 03'\, 37.86" \,E$.
c) River Thames, London, England, 11/06/2006, coordinates: 
$51^0 \, 28' \, 05.40" \,N$, $0^0\, 15'\, 18.12" \,E$.}
\end{figure*}



\begin{figure*}[!ht]
\centering
\includegraphics[width=16.5cm]{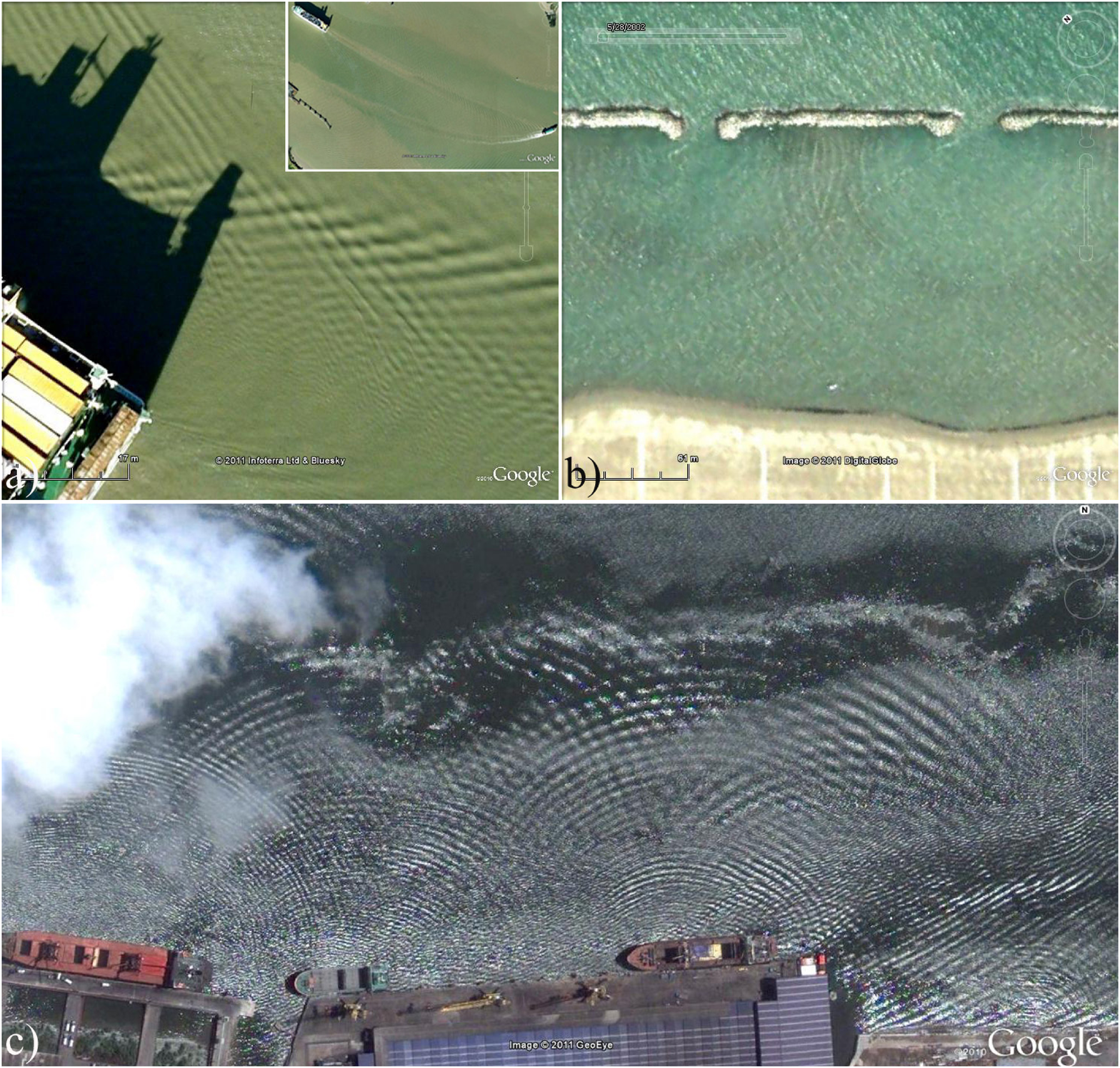}
\caption{
Interference between wave fronts producted by the wakes of two boats:
a) River Thames, London, England, 11/06/2006, coordinates: 
$51^0 \, 27' \, 40.79" \,N$, $0^0\, 16'\, 05.69" \,E$.
Interference between circular waves coming from two openings:
b) Rimini, Italy, 5/28/2002, coordinates: 
$44^0 \, 05' \, 15.02" \,N$, $12^0\, 32'\, 26.07" \,E$.
Interference  from secondary sources:
c) Chao Phraya River, Bangkok, Thailand, 8/20/2010, coordinates:
$13^0 \, 36' \, 40.11" \,N$, $100^0\, 34'\, 46.60" \,W$.}
\end{figure*}
 

For instance, in Fig. 1 we have  examples of diffraction that reminds us the one slit experiment: the protection barriers 
of  ports and beaches diffract approximate plane waves into circular waves.
The diffracted waves distribute their energy on circular wave fronts. It produces the circular erosions of the beach showed in Fig 1c. 
Fig 2a and Fig. 2b are other examples how waves can go beyond obstacles because of diffraction.

If waves reach an area where the water is shallower the friction with the bottom makes them slow down, see formula (3).
Waves in shallower water propagate more slowly and their wavelength decreases. They can change in direction and be subject 
to refraction. When wave fronts go forward to a straight shoreline at a certain angle, they bend and tend to become parallel 
to the line of the beach.  Examples of such  refraction is represented in Fig. 3.

A tool of Google Earth allows us to measure the distance among two geographical points. 
So students can measure the wavelength, and calculate the wave speed in deep water by applying Eq.  (\ref{Gw1}).
For instance, for $\lambda=100 \, m$ we have  $v_{phase}= 12.5 \, m/s$.
If it is possible estimate the depth of water, students can also calculate the speed in shallow water with Eq. (\ref{Gws}). 
For instance, with  $d= 1 \, m  $ we get  $v_{phase}= 3.1 \, m/s$.

When waves encounter an obstacle as a cliff or a wall, they reflect back upon themselves.     
In Fig. 4 there are examples of such reflections.

Examining Google Earth images students can verify the law of reflection: 
the angle of incidence of waves equals the angle of reflection.

Sometimes reflected waves interfere with coming waves, as we can observe in these pictures, 
(in shallow water also refraction and diffraction are often combined together, for instance, 
it probably happens in Fig. 2 and Fig. 3c).

The overlap of two crests or two troughs makes a crest or a trough even bigger. 
The overlap of a crest with a trough causes a cancellation of the wave perturbation. 
This is the interference between waves which we can see in Fig. 5.  
Fig. 5b may be an amusing example of two slit experiment.

\section{Conclusions}

Students can search out their own examples of wave phenomena on Google Earth. 
This is very instructive and entertaining for them. They, together with their teachers, 
can compare their own pictures with the traditional images in textbooks. 
Teachers could suggest looking for waves near big towns on the sea coast, big harbors, marina resorts and  big rivers.

Simple introductory books to the physics of water waves are \cite{Barber}, \cite{Bascom}. For other   less    
elementary introductions you could see, for example, \cite{Coulson}, \cite{Stoker}. General presentations about
waves are in \cite{Pierce}-\cite{Schiller}
 
\vskip.4cm

\noindent
{\Large{\bf Acknowledgments}}
\vskip.4cm
\noindent
I would like to thank  for useful comments and suggestions my friends  Beniamino Danese, Luca Chiari, Roberto Di Criscienzo, 
Onur Umucalilar and Lorenzo Sebastiani.

\vskip.4cm
 




\medskip



\begin{thebibliography}{99}
\bibitem{Feynman}
Feynman RP, Leighton RB \& Sands M 1989,  {\sl The Feynman Lectures on Physics} Vol. {\bf 1} (Reading MA: Addison-Wesley) pp 51/7-51/10.
	

\bibitem{Barber}
Barber NF  1969, {\sl Water Waves} (London: Wykeham   Publications). 



\bibitem{Bascom}
Bascom W  1964, {\sl Waves and Beaches}  (Garden City, New York: Anchor Books). 



\bibitem{Lighthill}
Lighthill J 2001, {\sl Waves in Fluids} (Cambridge: Cambridge University Press) pp 269-79.


\bibitem{Ryder}
Ryder BA, 2007 Journey to the Ends of the Earth {\sl Phys. Educ.} {\bf 42} pp 435-7.



\bibitem{Aguiar}
Aguiar CE and Souza AR  2009, Google Earth Physics  {\sl Phys. Educ.} {\bf 44} pp 624-626.


\bibitem{Logiuratoa}
Logiurato F and Danese B 2010, Physics of Waves with Google Earth  {\sl AIP Conf. Proc.} {\bf 1263} pp 208-211.



\bibitem{Logiuratob}
A shorter version than the present paper is  published in:
 
Logiurato F 2012, {\sl Phys. Educ.} {\bf 47}  pp 73-77.






\bibitem{Google}
http/earth.google.com/

\noindent
Google Earth is an interactive software that maps the Earth by satellite images. Its pictures are free if they are not used in commercial products or sold to others. Note that Google Earth Company periodically updates its    maps, so images change over time.  However, an option allows us to see also old pictures (clock icon on the upper toolbar).


						
\bibitem{French}
French AP 1971, {\sl Vibrations and Waves}  (London: Thomas Nelson and Sons LTD). 



\bibitem{Coulson}
Coulson CA 1977, {\sl Waves-A Mathematical Account of the Common Types of Wave Motion} (New York: Longman Inc).


\bibitem{Stoker}
Stoker  JJ 1957, {\sl Water Waves} (New York: Publishers LTD).


\bibitem{Pierce}
Pierce  JR 1981, {\sl Almost All About Waves} (The Massachusetts Institute of Technology).

\vskip.9cm
\rule{15cm}{0.5mm}

\newpage


\bibitem{Crowell}
Crowell  B 2006, {\sl Vibrations and Waves - Light and Matter 2.2 ed}, Vol. {\bf 3}, free physics textbooks at:
www.lightandmatter.com.


\bibitem{Giambattista}
Giambattista A, McCarthy Richardson  B and Richardson  RC 2010, {\sl College Physics} (New York: McGraw-Hill).


\bibitem{Schiller}
Schiller C 2011, {\sl Motion Mountain  - The Adventure of Physics 24.24 ed} Vol. {\bf 1},  free physics textbooks at:
www.motionmountain.net.



\end{thebibliography}
\end{document}